\documentclass[twocolumn]{aastex701}

\newcommand{\Msun}{$M_{\odot}$}
\newcommand{\HI}{\hbox{{\rm H}\kern 0.2em{\sc i}}}

\begin{document}

\title{A Naked Dwarf: Molecular Gas in the Completely Stripped {\HI} Tail of VCC~1249}

\author[orcid=0000-0002-3810-1806]{Bumhyun Lee}
\affiliation{Department of Astronomy, Yonsei University, 50 Yonsei-ro, Seodaemun-gu, Seoul 03722, Republic of Korea}
\affiliation{Korea Astronomy and Space Science Institute, 776 Daedeokdae-ro, Yuseong-gu, Daejeon 34055, Republic of Korea}
\email[show]{bhlee301@gmail.com}  

\author[orcid=0000-0003-1440-8552]{Aeree Chung}
\affiliation{Department of Astronomy, Yonsei University, 50 Yonsei-ro, Seodaemun-gu, Seoul 03722, Republic of Korea}
\email[]{achung@yonsei.ac.kr} 

\author[orcid=0000-0001-5965-252X]{Paolo Serra}
\affiliation{INAF - Osservatorio Astronomico di Cagliari, Via della Scienza 5, I-09047, Selargius (CA), Italy}
\email[]{paolo.serra@inaf.it} 


\author[orcid=0000-0001-7732-5338]{Nikki Zabel}
\affiliation{Department of Astronomy, University of Cape Town, Private Bag X3, Rondebosch 7701, South Africa}
\affiliation{South African Astronomical Observatory, PO Box 9, Observatory Cape Town 7935, South Africa}
\email[]{nikki.j.zabel@gmail.com} 

\author[orcid=0000-0002-5049-4390]{Sungsoon Lim}
\affiliation{Division of Science Education, Kangwon National University, Chuncheon 24341, Republic of Korea}
\affiliation{Kangwon National University Research Institute for Mathematical Sciences, Chuncheon 24341, Republic of Korea}
\email[]{slim@kangwon.ac.kr} 

\author[orcid=0000-0003-4048-2203]{Hyein Yoon}
\affiliation{Korea Astronomy and Space Science Institute, 776 Daedeokdae-ro, Yuseong-gu, Daejeon 34055, Republic of Korea}
\affiliation{Sydney Institute for Astronomy, School of Physics A28, University of Sydney, NSW 2006, Australia}
\email[]{hiyoon@kasi.re.kr} 

\author[orcid=0000-0002-9795-6433]{A. Boselli}
\altaffiliation{Scientific associate INAF - Osservatorio Astronomico di Cagliari, Via della Scienza 5, 09047 Selargius (CA), Italy}
\affiliation{Aix Marseille Univ, CNRS, CNES, LAM, Marseille, France}
\email[]{alessandro.boselli@lam.fr}

\author[orcid=0000-0002-9043-8764]{Matteo Fossati}
\affiliation{
Dipartimento di Fisica ``G. Occhialini'', 
Universit\`a degli Studi di Milano-Bicocca, 
Piazza della Scienza 3, I-20126 Milano, Italy}
\affiliation{INAF – Osservatorio Astronomico di Brera, 
Via Brera 28,
I-21021 Milano, Italy}
\email[]{matteo.fossati@unimib.it}

\author[orcid=0000-0003-1647-3286]{Yongjung Kim}
\affiliation{School of Liberal Studies, Sejong University, 209 Neungdong-ro, Gwangjin-gu, Seoul 05006, Republic of Korea}
\affiliation{Department of Physics and Astronomy, Sejong University, 209 Neungdong-ro, Gwangjin-gu, Seoul 05006, Republic of Korea}
\affiliation{Department of Astronomy and Space Science, Sejong University, 209 Neungdong-ro, Gwangjin-gu, Seoul 05006, Republic of Korea}
\email{yjkim.ast@gmail.com}

\author[orcid=0000-0003-2475-7983,gname=Tomonari,sname=Michiyama]{Tomonari Michiyama}
\affiliation{Faculty of Information Science, Shunan University, 843-4-2 Gakuendai, Shunan, Yamaguchi, 745-8566, Japan}
\email[]{t.michiyama.astr@gmail.com} 

\author[orcid=0000-0002-8136-8127]{Juan Molina}
\affiliation{Instituto de F\'isica y Astronom\'ia, Universidad de Valpara\'iso, Avda. Gran Breta\~na 1111, Valpara\'iso, Chile}
\affiliation{Millenium Nucleus for Galaxies (MINGAL)}
\email[]{juan.molinato@uv.cl} 

\author[orcid=0000-0003-3932-0952]{Kana Morokuma-Matsui}
\affiliation{Institute of Astronomy, Graduate School of Science, The University of Tokyo, 2-21-1, Osawa, Mitaka, Tokyo 181-0015, Japan}
\email[]{kanamoro@ioa.s.u-tokyo.ac.jp} 

\author[orcid=0000-0002-6810-1778]{Jaehyun Lee}
\affiliation{Korea Astronomy and Space Science Institute, 776 Daedeokdae-ro, Yuseong-gu, Daejeon 34055, Republic of Korea}
\email[]{jaehyun@kasi.re.kr} 

\author[orcid=0000-0003-3301-759X]{Jeong Hwan Lee}
\affiliation{Research Institute of Basic Sciences, Seoul National University, Seoul 08826, Republic of Korea}
\affiliation{Department of Physics and Astronomy, Seoul National University, 1 Gwanak-ro, Gwanak-gu, Seoul 08826, Republic of Korea}
\email[]{joungh93@gmail.com} 

\author[orcid=0000-0002-8379-0604]{Se-Heon Oh}
\affiliation{Department of Physics and Astronomy, Sejong University, 209 Neungdong-ro, Gwangjin-gu, Seoul 05006, Republic of Korea}
\email[]{seheon.oh@sejong.ac.kr} 


\begin{abstract}
We present the first observational hints of the severe removal of both molecular and {\HI} gas from the dwarf galaxy VCC~1249. This extreme stripping event is thought to be driven by the combined effects of tidal interaction and ram pressure. Using deep CO (2$-$1) observations from the James Clerk Maxwell Telescope (JCMT), we obtained marginal CO detections in three regions within the stripped {\HI} tail, with molecular masses of $\sim$10$^{5}$ to 10$^{6}${\Msun}, comparable to typical masses of giant molecular clouds.  In contrast, we did not find CO emission within the stellar disk of VCC~1249. This indicates the severe removal of cold gas, which likely caused the sudden cessation of star formation in the galaxy. This identifies VCC~1249 as a unique laboratory for witnessing the rapid, environmentally-driven quenching of a dwarf galaxy. Our findings provide a critical observational link between gas removal mechanisms and the dramatic phase transition of cluster dwarfs from star-forming to quiescent systems.
\end{abstract}






\section{Introduction} 
Dwarf galaxies are the most dominant population in nearby clusters, comprising $70-90\%$ of all member galaxies \citep{choque-challapa2021}. Their shallow gravitational potential makes them highly sensitive to environmental processes, such as tidal interaction and ram pressure stripping (RPS) \citep[see][for a review]{boselli2014,cortese2021}. These mechanisms can rapidly quench star formation by removing cold interstellar medium (ISM; molecular and {\HI} gas). This is supported by Fornax blue dwarfs losing {\HI} gas faster than their color changes \citep{kleiner2023}, and by a truncated {\HI} mass function \citep{kleiner2025}.

While studying the molecular gas, which is direct fuel for star formation, of dwarf galaxies in the cluster environment is necessary for obtaining a complete picture of galaxy evolution in dense regions, previous observational studies have mostly focused on non-dwarf populations \citep[e.g.,][]{lee2017,brown2021,moretti2026}. To date, there are only a handful of studies that have specifically probed environmental effects on the molecular gas of dwarf galaxies \citep[e.g.,][]{jachym2013,lisenfeld2016,grossi2016,zabel2024}. In particular, \cite{zabel2019} found that some dwarf galaxies show molecular gas stripping in the Fornax cluster.

\begin{figure*}[!htbp]
\begin{center}
\includegraphics[width=1.0\textwidth]{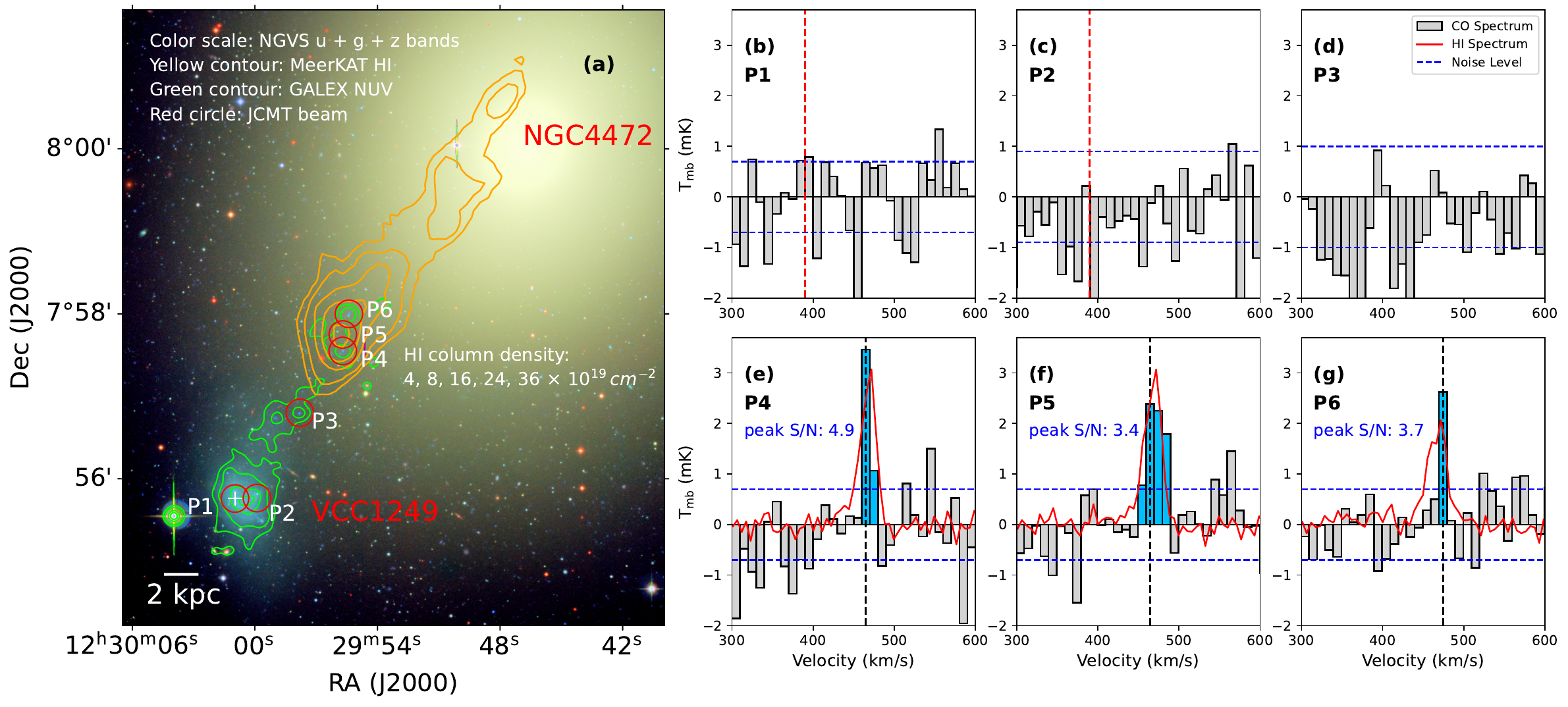}
\caption{(a): A panchromatic map of VCC~1249, including optical color image created by combining the Next Generation Virgo Cluster Survey (NGVS) $u$, $g$, $z$ band data \citep{ferrarese2012,ferrarese2016}, GALEX NUV data (green contours), and MeerKAT {\HI} data (yellow contours, \citealt{degasperin2025}, Serra et al. in prep.). The GALEX NUV data, based on GALEX GR6 data from the MAST GALEX archive, were retrieved through NASA's SkyView facility \citep{mcglynn1998,mast2013}. Red circles indicate the JCMT CO observing fields and the JCMT beam size ($\sim$20$\arcsec$). (b)-(g): CO spectra in six regions of VCC~1249. The observed CO spectra ($\Delta v_{\rm CO}$ = 10 km~s$^{-1}$) are shown in gray. CO detections are shown in blue. Black vertical lines indicate the peak CO velocities. Red vertical lines show the system velocity ($\sim$390~km~s$^{-1}$; \citealt{arrigoni_battaia2012}) of VCC~1249. The MeerKAT {\HI} spectra ($\Delta v_{\rm HI}$ = 5.5 km~s$^{-1}$) within the JCMT observing fields are shown in red.} \label{fig:image_spec}
\end{center}
\end{figure*}

VCC~1249 is an interesting dwarf galaxy (see physical properties in Table~\ref{tab:basic}) in the Virgo cluster that appears to be undergoing both tidal interaction and RPS. Given that the projected distance between VCC~1249 and NGC~4472 (M49, a giant elliptical galaxy, $M_\star$: $8\times10^{11}${\Msun}, located at the core of the Virgo B subcluster; \citealt{cote2003,mihos2013}) is $\sim$27 kpc and that the stellar disk is distorted together with faint stellar tail structures \citep[][see Figure~2 of \citealt{arrigoni_battaia2012}]{lee1997}, VCC~1249 is likely gravitationally interacting with NGC~4472. In addition, RPS by the hot halo gas of NGC~4472 \citep{irwin1996} is likely removing the {\HI} gas of VCC~1249. {\HI} observations from Arecibo, Very Large Array, and MeerKAT reveal a long {\HI} tail ($M_{\rm H\kern 0.2em{\sc I}} = 4.7–7.9 \times 10^{7}~M_\odot$) that is entirely detached from the stellar disk, with no {\HI} emission detected within the disk \citep[see Figure~\ref{fig:image_spec}(a);][Serra et al. in prep.]{sancisi1987, mcnamara1994, degasperin2025}. The morphology of stripped {\HI} gas closely resembles the {\HI} morphology of other RPS cases \citep[e.g.,][]{chung2007,ramatsoku2025}. The side directly affected by ram pressure has high gas density with narrow contour spacing, while the opposite side shows a long tail (Figure~\ref{fig:image_spec}). These features support that VCC~1249 is undergoing active RPS. As a likely consequence of these combined environmental processes, recent star-forming regions, such as UV and H$\alpha$ knots, are found outside the stellar disk (see Figure~\ref{fig:image_spec}(a); \citealt{lee1997,arrigoni_battaia2012}). These characteristics make VCC~1249 an ideal target to test whether environmental effects can completely remove molecular gas alongside the {\HI} gas.

In this letter, we present new James Clerk Maxwell Telescope (JCMT) $^{12}$CO(2$-$1) data of VCC~1249. Notably, we provide a new observational snapshot of a critical evolutionary stage where a dwarf galaxy has experienced a severe  loss of both molecular and {\HI} gas from its stellar disk. What distinguishes this system from gas-free and passive dwarf galaxies is that the removed gas remains detectable around VCC~1249, indicating a critical transitional phase. We adopt a distance of 16.9 Mpc to NGC~4472 \citep{mei2007}. We assume that VCC~1249 is at the same distance of NGC~4472.

\begin{figure}
\center
\includegraphics[scale=0.45]{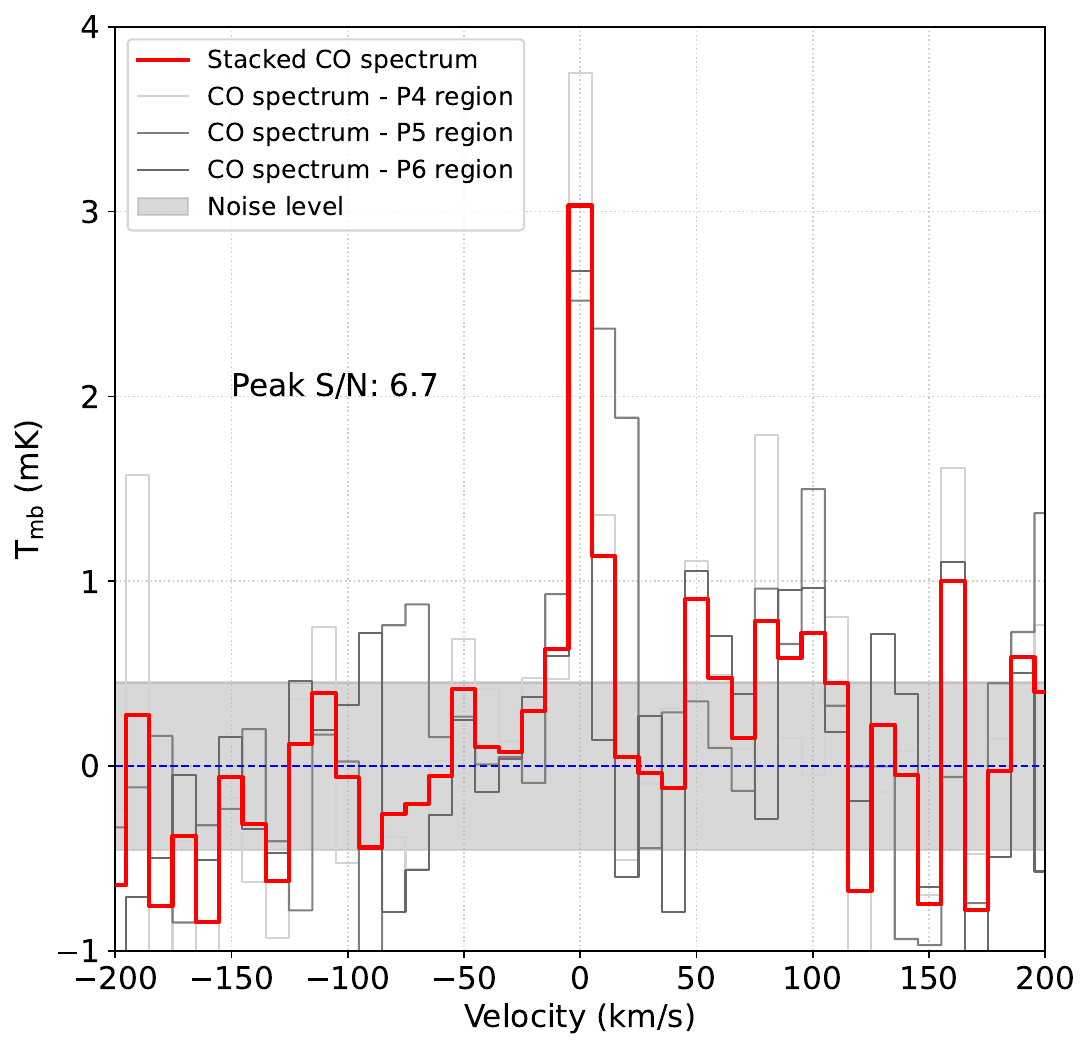}
\caption{Stacked CO spectrum (red) by combining CO spectra from P4 - P6 regions. The peak S/N is $\sim$6.7.} \label{fig:stack}
\end{figure}

\section{JCMT observations and data reduction} \label{sec:obs}
Our JCMT observations were carried out from December 2024 to April 2025 through three JCMT observing programs (Program ID: E24XP015, R25AP005, E25AK004; PI: B. Lee). Using the $^\backprime\bar{\rm U}^\backprime\bar{\rm u}$ receiver, we observed $^{12}$CO (2$-$1) ($\nu_{\rm rest} = 230.538~$GHz) in six regions of VCC~1249, of which two regions (P1 and P2) within the stellar disk and the other four regions (P3 - P6) outside of stellar disk (Figure~\ref{fig:image_spec}(a)). The beam size is 20$\arcsec$, corresponding to $\sim$1.64 kpc. The total on-source time was 37 hours, with individual regions ranging from 4.0$-$7.5 hours (see Table~\ref{tab:co_result}). The JCMT CO data were reduced using the Starlink software \citep{currie2014}. We binned the CO spectrum from a velocity resolution of $\sim$1.3~km~s$^{-1}$ to 10~km~s$^{-1}$. We converted the corrected antenna temperature ($T^*_{\rm A}$) to the main beam temperature ($T_{\rm mb}$) using $T_{\rm mb} = T^*_{\rm A}$/$B\rm_{eff}$, where the beam efficiency ($B\rm_{eff}$) is 0.66 for $^\backprime\bar{\rm U}^\backprime\bar{\rm u}$. We used the $T_{\rm mb}$ scale for further analysis. The rms noise levels are $\sim$0.7$-$1.0 mK ($T_{\rm mb}$).

\begin{deluxetable}{l @{\hspace{2.5cm}} c}[h]
\tablecaption{Basic information of VCC~1249 \label{tab:basic}}
\tablehead{\colhead{Parameter} & \colhead{Value}}

\startdata
Name & VCC~1249; UGC~7636 \\
R.A. (J2000)$^{a}$ & $12^h30^m01^s$ \\
Decl. (J2000)$^{a}$ & $+07\degr55\arcmin46\arcsec$ \\
Morphological type$^{a}$    & dIrr \\
Inclination ($\degr$)$^{b}$ & 53.8 \\
Position angle ($\degr$)$^{b}$ & 5.2 \\
$D_{25}$ (arcmin; kpc)$^{b}$ & 1.0; 4.9 \\
$v_{\rm opt}$ (km~s$^{-1}$)$^{a}$ & 390 \\
$M_{\rm stellar}$~(\Msun)$^{a}$ & 1.2 $\times$ 10$^9$ \\
$M_{\rm\tiny {\HI}}$~(\Msun)$^{c}$ & 7.9 $\times$ 10$^7$ \\
\enddata
\tablecomments{(a) \cite{arrigoni_battaia2012}; (b) HyperLeda \citep[][\url{http://leda.univ-lyon1.fr/}]{makarov2014}; (c) \HI\ mass estimated using the \HI\ data from the MeerKAT survey of the Virgo cluster \citep[][Serra et al. in prep.]{degasperin2025}.}
\end{deluxetable}

\section{Results} \label{sec:result}
We determine CO detection based on two criteria: (i) the peak CO emission has a signal-to-noise ratio (S/N) $>$3, and (ii) the CO peak velocity is approximately consistent with the {\HI} peak velocity. After confirming the peak emission, consecutive neighboring CO emission channels with the S/N $>$1 are also included in the detection. 

We identify marginal CO detections in P4 - P6 regions of the {\HI} tail with peak S/N $\sim$3.4$-$4.9 (Figure~\ref{fig:image_spec}(e)-(g)). In addition, their integrated S/N values range from 3.7 to 5.2 (Table~\ref{tab:co_result}). In particular, within the same JCMT observing fields, the peak velocities ($\sim$472~km~s$^{-1}$) of the {\HI} profiles obtained from the MeerKAT data \citep{degasperin2025} show good agreement with those ($\sim$465~km~s$^{-1}$ for P4 and P5 regions and $\sim$474~km~s$^{-1}$ for P6 region) of the JCMT CO profiles. Although low S/N requires us to classify these features conservatively, the kinematic coincidence between {\HI} and CO profiles supports that the marginal CO features may be associated with the stripped {\HI} gas in the tail.

To enhance the S/N of the CO detection, we stacked the CO spectra from P4 - P6 regions after shifting the CO peak velocities to zero. The stacked CO spectrum with peak S/N $\sim$6.7 suggests the presence of CO emission in the {\HI} tail region of VCC~1249 (Figure~\ref{fig:stack}). Previous CO observation with the NRAO 12m single-dish telescope reported non-detections in the {\HI} peak position (i.e., P5 region) \citep{irwin1997} due to the sensitivity limit. Its 3$\sigma$ upper limit is 9.6$\times$10$^{6}${\Msun} using the Milky Way (MW) conversion factor. However, our deep JCMT data provide the first observational evidence of CO emission in the {\HI} tail of VCC~1249, although individual features are marginal. On the other hand, we did not detect any CO emission within the stellar disk (i.e., P1 and P2) or in one bright NUV region (P3) between the stellar disk and the {\HI} tail (Figure~\ref{fig:image_spec}(a)-(c))). These results suggest that the molecular gas of VCC~1249 appears to have been severely removed from its stellar disk together with its {\HI} gas by the environmental effects. 

For CO-detected regions (P4 - P6), we calculated integrated CO intensities, luminosities, and molecular gas masses (Table~\ref{tab:co_result}). The CO luminosities are estimated using the following equation \citep{solomon2005}:

\begin{equation}
L^{'}_{\rm CO}=3.25\times10^7 S_{\rm CO}~\nu_{\rm obs}^{-2}~D_{\rm L}^2~(1+\textit{z})^{-3} 
\label{eqn:colum}
\end{equation}

\noindent in K~km~s$^{-1}$~pc$^{2}$, where $S_{\rm CO}$ is the integrated CO flux in Jy~km~s$^{-1}$, $D_{\rm L}$ is the luminosity distance in Mpc, $\nu_{\rm obs}$ is the observing frequency in GHz, and \textit{z} is the redshift. We calculated the molecular gas mass with the equation: $M_{\rm H_{2}}=\frac{\alpha_{\rm CO}}{R_{21}} L^{'}_{\rm CO}$. Since VCC~1249 is a low metallicity system, we used two different CO-to-H$_{2}$ conversion factors: (i) the MW conversion factor ($\alpha_{\rm CO}$~=~4.35~{\Msun}~pc$^{-2}$~(K~km~s$^{-1}$)$^{-1}$; \citealt{bolatto2013}), and (ii) the metallicity dependent conversion factor calculated by following the Equation (25) of \cite{accurso2017} and the approach of \cite{zabel2019} ($\alpha_{\rm CO}$~=~6.05~ {\Msun}~pc$^{-2}$~(K~km~s$^{-1}$)$^{-1}$ for 12+log(O/H)$\approx$8.35, which is derived by the mass-metallicity relation; \citealt{arrigoni_battaia2012}). We also adopted a typical CO (2--1)/(1--0) ratio ($R_{21}\approx0.7$; \citealt{leroy2013}).

The molecular gas masses for P4 - P6 regions are $\sim$$10^{5}-10^{6}${\Msun}, comparable to the typical mass of giant molecular clouds (GMCs). Since the faint CO emission in three regions appears in only a few channels, it is difficult to accurately estimate their linewidths. 

For non-detections of CO emission, we derived 3$\sigma$ upper limits of the molecular gas mass, assuming CO linewidths of 50~km~s$^{-1}$ for P1 and P2 regions and 10~km~s$^{-1}$ for P3 region (Table~\ref {tab:co_result}).

\begin{deluxetable*}{ccccccccc}
\tablecaption{Results of the JCMT observations of VCC 1249 \label{tab:co_result}}
\tablehead{
\colhead{Region} & \colhead{R.A.} & \colhead{Decl.} & \colhead{on-source time} & \colhead{rms noise} & \colhead{$I_{\rm CO}$} & \colhead{log $L_{\rm CO}^{'}$} & \colhead{log $M_{\rm H2, MW}$} & \colhead{log $M_{\rm H2, met}$}
\\
\colhead{} & \colhead{(h:m:s)} & \colhead{(d:m:s)} & \colhead{(hours)} & \colhead{(mK)} & \colhead{(K~km~s$^{-1}$)} & \colhead{(K~km~s$^{-1}$~pc$^{2}$)} & \colhead{(\Msun)} & \colhead{(\Msun)}
\\
\colhead{(1)} & \colhead{(2)} & \colhead{(3)} & \colhead{(4)} & \colhead{(5)} & \colhead{(6)} & \colhead{(7)} & \colhead{(8)} & \colhead{(9)}
}
\startdata
P1 & 12:30:01.0 & +07:55:46 & 5.5 & 0.7 & $<$0.071 & $<$5.33 & $<$6.13 & $<$6.27 \\
P2 & 12:29:59.9 & +07:55:46 & 4.0 & 0.9 & $<$0.091 & $<$5.44 & $<$6.24 & $<$6.38 \\
P3 & 12:29:57.8 & +07:56:48 & 5.0 & 1.0 & $<$0.045 & $<$5.14 & $<$5.93 & $<$6.08 \\
P4 & 12:29:55.7 & +07:57:33 & 7.5 & 0.7 & 0.045$\pm$0.010 & 5.14$\pm$0.10 & 5.93$\pm$0.10 & 6.08$\pm$0.10 \\
P5 & 12:29:55.7 & +07:57:45 & 7.5 & 0.7 & 0.073$\pm$0.014 & 5.34$\pm$0.08 & 6.14$\pm$0.08 & 6.28$\pm$0.08 \\
P6 & 12:29:55.4 & +07:58:00 & 7.5 & 0.7 & 0.026$\pm$0.007 & 4.89$\pm$0.12 & 5.69$\pm$0.12 & 5.83$\pm$0.12 \\
\enddata
\tablecomments{(1) region of VCC~1249; (2) R.A. (J2000); (3) Decl. (J2000); (4) on-source time; (5) rms noise at a velocity resolution of 10~km~s$^{-1}$; (6) velocity-integrated CO intensity; (7) CO luminosity; (8) \& (9) molecular gas mass derived using the MW conversion factor ($\alpha_{\rm CO}$~=~4.35 {\Msun}~pc$^{-2}$~(K~km~s$^{-1}$)$^{-1}$) and the metallicity-dependent conversion factor ($\alpha_{\rm CO}$~=~6.05 {\Msun}~pc$^{-2}$~(K~km~s$^{-1}$)$^{-1}$ for 12+log(O/H)$\approx$8.35; \citealt{arrigoni_battaia2012}).}
\end{deluxetable*}

\section{Discussion} \label{sec:discussion}
\subsection{Can CO-dark gas account for the non-detection of CO in the disk?} \label{subsec:co-dark}
If VCC~1249 contains a large fraction of CO-dark molecular gas because of its low metallicity \citep{wolfire2010}, this could explain the non-detection of CO within its stellar disk. If CO-dark molecular gas is surviving there, ongoing star formation would be expected. However, this is not found, as indicated by the absence of H$\alpha$ emission within the stellar disk \citep{arrigoni_battaia2012}. Furthermore, we do not attribute the non-detection of CO in the stellar disk solely to CO-dark molecular gas for the following reasons. (i) Some Virgo dwarfs with stellar masses and metallicities comparable to those of VCC~1249 show clear CO detections in their stellar disks (e.g., VCC~340, \citealt{grossi2016}). (ii) the CO detection rate remains relatively high for galaxies with 12+log(O/H) $>$ 8.2 \citep{leroy2005}, suggesting that the fraction of CO-dark molecular gas in VCC 1249 is likely lower than in more metal-poor dwarfs. (iii) We obtained marginal CO detections in the stripped gas region, where the metallicity is expected to be similar to that of the stellar disk \citep{arrigoni_battaia2012}. Taken together, these factors suggest that if molecular gas were present in the central region of VCC 1249, we would expect to detect CO emission. Consequently, this interpretation supports the severe removal of molecular gas from VCC~1249.

\subsection{Environmental effects on the removal of cold gas components from VCC~1249}
 \label{subsec:removal}
Since VCC~1249 appears to be undergoing both tidal interaction and RPS \citep{mcnamara1994,arrigoni_battaia2012}, we examine whether these environmental processes can remove its {\HI} and H$_{2}$ components. Severe tidal interaction is expected between VCC~1249 and NGC~4472, given their close 27~kpc projected separation. 

We estimated the tidal radius of VCC~1249 using the following relation \citep{johnson2015}:

\begin{equation}
r_{\rm T} \approx R(m/(M(3+e)))^{1/3}, 
\label{eqn:ram}
\end{equation}

\noindent where $m$ and $M$ are stellar masses of smaller (VCC~1249) and larger (NGC~4472) galaxies in a two-body system, $R$ is their projected separation, and $e$ is the orbital eccentricity of VCC~1249. Considering the full range of possible eccentricities (0 $<$ e $<$ 1), the resulting tidal radius varies between 1.9 and 2.1 kpc. These values lie within the stellar disk of VCC~1249 (radius: $\sim$2.45 kpc). Furthermore, at its 15-kpc closest approach to NGC~4472, the tidal radius of VCC~1249 likely reached $\sim$1.2~kpc for e$=$0.5, suggesting that tidal force potentially affected its inner structure. These estimates suggest that gas and stars outside the tidal radius are highly vulnerable to tidal stripping. Indeed, VCC~1249 exhibits faint stellar tails \citep{arrigoni_battaia2012}. Moreover, the tidal stripping may perturb the gravitational potential of VCC~1249 and/or displace gas to larger radii \citep[e.g.,][]{boselli2022,lin2023}. If this occurs, the local anchoring pressure (i.e., restoring force per unit area) would be reduced, making the cold gas components even more susceptible to RPS \citep{mcnamara1994,arrigoni_battaia2012}.

\begin{figure}
\center
\includegraphics[scale=0.45]{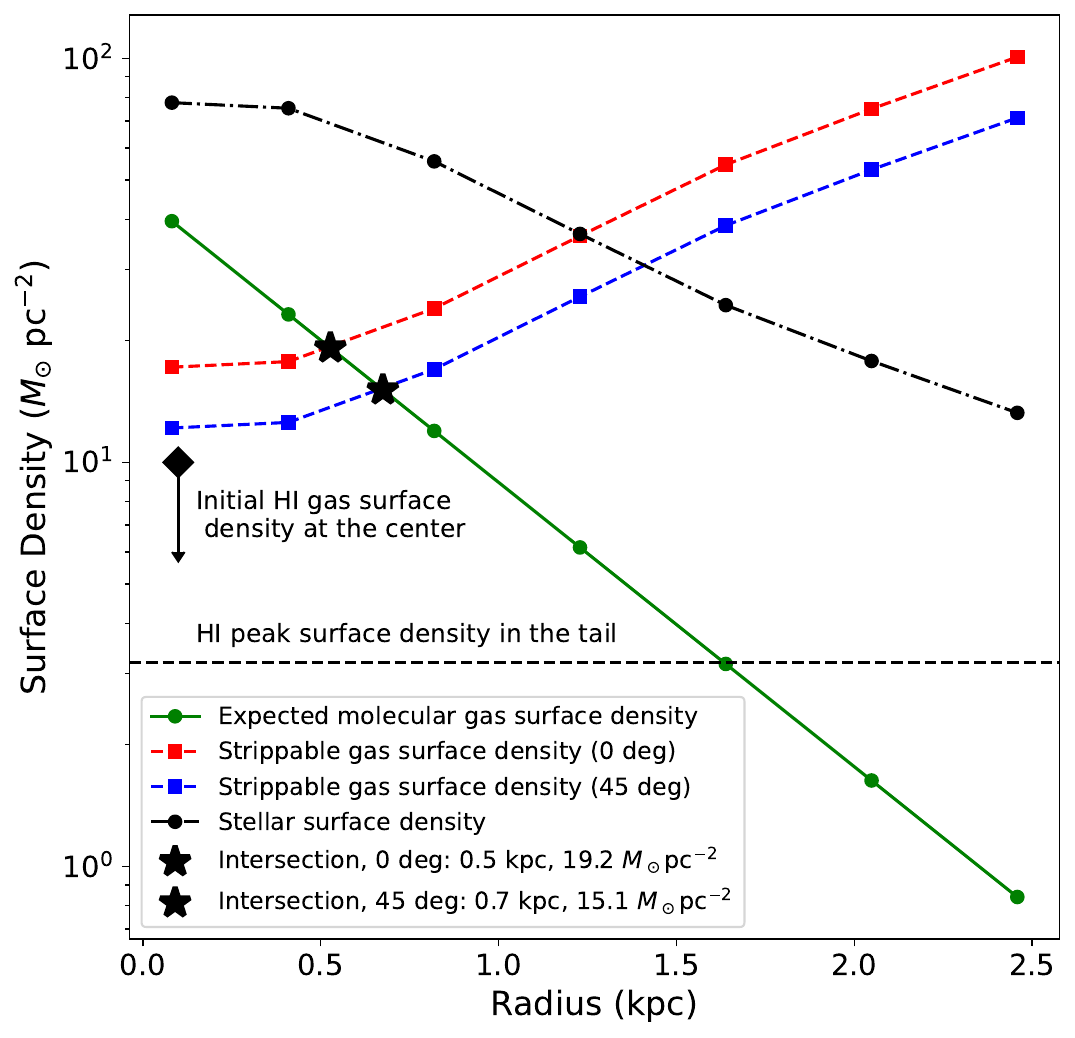}
\caption{Strippable gas surface densities due to RPS as a function of radius for encounter angles of 0 and 45 degrees (red and blue dashed lines), alongside the expected molecular gas profile (green line) and the stellar surface density profile (black line) of VCC~1249. Black stars indicate strippable molecular gas densities (19.2~{\Msun}~$\rm pc^{-2}$ and 15.1~{\Msun}~$\rm pc^{-2}$) at radii of 0.5 kpc and 0.7 kpc, respectively. Black horizontal line and black diamond indicate the peak density of the {\HI} tail ($\sim$3.2~{\Msun}~$\rm pc^{-2}$) and the expected {\HI} density in the center ($<$10~{\Msun}~$\rm pc^{-2}$), respectively.} \label{fig:rps_cal}
\end{figure}

In addition, we probed the impact of RPS on VCC~1249. If the ram pressure exceeds the gravitational anchoring pressure, gas can be stripped from the stellar disk \citep[][]{gunn1972}. As VCC~1249 is experiencing active RPS due to its interaction with the hot halo of NGC~4472, we calculated the ram pressure using the equation \citep{gunn1972,wang2021,lees2022}:

\begin{equation}
P_{\rm ram}= \rho_{\rm halo} \Delta v^2 cos^{2}{\theta}, 
\label{eqn:ram}
\end{equation}

\noindent where $\rho_{\rm halo}$ is the hot gas density ($\rho_{\rm halo}=1.4\times m_{\rm p}\times n_{{\rm halo}}$). The factor 1.4 accounts for the mass contribution from helium, $m_{\rm p}$ is the proton mass, and $\Delta v$ is the velocity difference (611 km~s$^{-1}$, \citealt{arrigoni_battaia2012}) between VCC~1249 and NGC~4472. Using the ROSAT X-ray measurements and the beta model \citep{schindler1999,vollmer2009}, we derived a hydrogen number density of $n_{{\rm halo}}$$\sim$2.8$\times$10$^{-3}$~cm$^{-3}$ for the NGC~4472 halo gas at the VCC~1249 location. The encounter angle ($\theta$) between the ram pressure wind and the stellar disk is also taken into account \citep{roediger2006,lee2022}. In this work, we adopted two encounter angles of 0 (face-on stripping) and 45 degrees.

The anchoring pressure is estimated as follows \citep[][]{lees2022}:

\begin{equation}
P_{\rm anchor} = 2 \pi G \Sigma_* \Sigma_{{\rm gas}} , 
\label{eqn:anchor_pressure}
\end{equation}

\noindent where $G$ is the gravitational constant, and $\Sigma_*$ and $\Sigma_{{\rm gas}}$ are stellar and gas ({\HI} or molecular gas) surface densities, respectively. We calculated the anchoring pressure as a function of galactic radius using the stellar surface density radial profile derived from the NGVS data, together with the gas surface density. By comparing the ram pressure with the anchoring pressure, we derived the threshold gas surface density for strippable gas ($\Sigma_{\rm gas, thres} < P_{\rm ram} / 2 \pi G \Sigma_*$). If the gas surface density is lower than this threshold at a given radius, ram pressure can remove the gas from VCC~1249. Figure~\ref{fig:rps_cal} shows the strippable gas surface density as a function of galactic radius. It is worth noting that the anchoring pressure calculated using Eq.~\ref{eqn:anchor_pressure} under a thin-disk assumption \citep[e.g.,][]{choi2022} may be overestimated, compared to the true value in irregular dwarf galaxies, due to their asymmetric and complex structure.

For {\HI} gas, we considered two cases: (i) the peak surface density of the stripped {\HI} tail, $\sim$3.2~{\Msun}~$\rm pc^{-2}$ (based on MeerKAT data), and (ii) an assumption of initial {\HI} gas surface density in the central region, $<$10~{\Msun}~$\rm pc^{-2}$ \citep{wang2016}. In both cases, the estimated ram pressure is sufficient to strip the {\HI} gas, leading to the complete removal of {\HI} gas from the stellar disk (Figure~\ref{fig:rps_cal}).

For molecular gas, assuming an exponential gas disk, we derived the expected molecular gas surface density profile using the expected molecular gas mass (1.08$\times$10$^{8}$ {\Msun}), calculated from the scaling relation between H$_{2}$ mass and stellar mass, and the expected $R_{90,\rm mol}$ (the radius containing 90\% of the flux, 2.4 kpc) \citep{cicone2017,brown2021}. By comparing this expected molecular gas surface density profile with the threshold for strippable gas, we determined the stripping radii and their corresponding gas surface densities. As shown in Figure~\ref{fig:rps_cal}, given the assumed molecular gas radial profile, stripping is possible down to radii of 0.5 and 0.7 kpc for encounter angles of 0 and 45 degrees, respectively. These radii correspond to maximum strippable gas surface densities of 19.2 and 15.1 {\Msun}~$\rm pc^{-2}$, respectively. However, the current ram pressure is not strong enough to remove higher-density gas in the innermost region. As discussed, the tidal interaction may enhance the RPS efficiency \citep{gavazzi2001,boselli2022,lin2023}. Therefore, the combined effect of ram pressure and tidal interaction may lead to the severe stripping of molecular gas from VCC~1249 \citep{lee1997,lee2000,mcnamara1994,arrigoni_battaia2012}. However, we note that the projection effect on the separation and velocity difference causes some uncertainties in our interpretation, as the projected values may deviate from the true 3D physical separation and relative velocity.

\subsection{Origin of molecular gas in the {\HI} tail} \label{sec:origin}
Recent observations have revealed molecular clouds in the tails of galaxies undergoing strong RPS \citep[e.g.,][]{lee2018,moretti2018,moretti2026,jachym2014,jachym2017,jachym2019,jachym2022}. Based on these studies, and on the fact that the estimated molecular gas masses are comparable to those of GMCs, we treat the molecular gas detected in the tail as GMC-like molecular structures in the following discussion. These previous studies suggest two possible scenarios for the origin of extraplanar molecular clouds: (i) direct stripping of molecular clouds from the stellar disk, or (ii) in situ formation from stripped atomic gas.

By following the same approach of \cite{arrigoni_battaia2012}, we calculated the travel time for the stripped gas to reach its current position from the stellar disk. Given the projected relative velocity of $\sim$65 km~s$^{-1}$ between VCC~1249 and the {\HI} peak (i.e., P5 region) in the {\HI} tail, and a projected separation of $\sim$11.7 kpc, the travel time is $\sim$176 Myr (cf. $\sim$124 Myr from \cite{arrigoni_battaia2012}, by adopting a 10 kpc distance and the relative velocity of 79 km~s$^{-1}$). Considering typical timescales for the transition of {\HI} to H$_{2}$ and the formation of molecular clouds ($\sim$10 Myr each; \citealt{chevance2023}), 176 Myr is sufficient time for molecular clouds to form in situ within the {\HI} tail. Conversely, if molecular clouds were directly stripped from the disk, they would be expected to dissipate before reaching the current {\HI} tail location, as typical GMC lifetimes are only 10$-$30 Myr \citep{chevance2023}, shorter than the 176 Myr travel time.

Based on these comparisons, the in-situ formation scenario appears more favorable. Furthermore, recent numerical simulations support in-situ formation in the far tail \citep{tonnesen2021,lee2022}. Specifically, \cite{lee2022} demonstrated that while the direct stripping dominates near the stellar disk ($d<10$ kpc), the in-situ formation becomes the primary mechanism in the more distant regions of the tail. As a large amount of ISM is stripped and mixes with the hot gas, it leads to the formation of abundant warm ionized clouds that are capable of cooling and collapsing within 100 Myr. This process facilitates the in-situ formation of molecular clouds.

\subsection{Implications of complete removal of cold gas} \label{subsec:implication}
The significant removal of both molecular and {\HI} gas suggests that star formation within the stellar disk of VCC~1249 was quenched abruptly. While there is no H$\alpha$ emission in the stellar disk of VCC~1249, FUV emission is found there. This suggests that the last star formation activity occurred $\sim$100 Myr ago and there was no recent star formation within the last 20 Myr \citep{arrigoni_battaia2012}. In addition, \cite{arrigoni_battaia2012} reported a sudden cessation of star formation 200 Myr ago in VCC~1249, using the spectral energy distribution fitting analysis. This timescale is roughly consistent with the estimated 175 Myr travel time of the stripped {\HI} gas. Consequently, the severe removal of the cold ISM can drive a rapid transition from a gas-rich, star-forming dwarf to a passive, quiescent system. This case demonstrates that dwarf galaxies in dense environments can evolve rapidly due to extreme environmental effects \citep{boselli2008,junais2022}. 

The expected {\HI} mass of VCC~1249 is 1.7$\times$10$^{9}$~{\Msun}, calculated using the scaling relation between {\HI} mass and stellar mass \citep{parkash2018}. The current {\HI} mass in the tail is only 7.9$\times$10$^{7}$~{\Msun}, representing 5\% of the expected {\HI} mass. Considering the expected molecular gas mass (1.1$\times$10$^{8}$~{\Msun}), Only 2.5\% of this expected molecular gas mass remains in the tail now, corresponding to 2.7$\times$10$^{6}$~{\Msun}. These results imply that a significant fraction of cold gas ({\HI} and molecular gas) has been removed from the stellar disk and mixed into the NGC~4472 halo. Continued interaction with the intracluster medium (ICM) in the Virgo cluster may further mix this stripped ISM with the ICM. Consequently, these processes likely contribute to chemical enrichment of both the hot halo of NGC~4472 and the Virgo ICM \citep[e.g.,][]{domainko2006,boselli2022}. 

High-resolution observations with the Atacama Large Millimeter/submillimeter Array (ALMA) are required to probe whether the detected molecular gas exists in the form of molecular clouds and to characterize their detailed properties (e.g., distribution and size). Deeper ALMA follow-up will also help confirm if molecular clouds survive within VCC~1249, as the current JCMT non-detection may be sensitivity-limited.


\begin{acknowledgments}
B. L. was supported by the National Research Foundation of Korea (NRF) grant funded by the Korea goverment (MSIT) (RS-2026-25498235). A. C. acknowledges support by the NRF, grant Nos. RS-2022-NR069020 and RS-2022-NR070872. The INAF - OAC computer cluster used in this work has been acquired within a project aimed to enhance the Sardinia Radio Telescope (SRT). The Enhancement of the SRT for the study of the Universe at high radio frequencies is financially supported by the National Operative Program (Programma Operativo Nazionale - PON) of the Italian Ministry of University and Research "Research and Innovation 2014-2020", Notice D.D. 424 of 28/02/2018 for the granting of funding aimed at strengthening research infrastructures, in implementation of the Action II.1 – Project Proposal PIR01\_00010. Co-funded by the European Union (ERC, ULU, 101086378). NZ is supported through the South African Research Chairs Initiative of the Department of Science and Technology and National Research Foundation. S. L. was supported by the National Research Foundation of Korea (NRF) grant funded by the Korea government (MSIT, No. 2710105275). H. Y. was supported by the National Research Foundation of Korea (NRF) grant funded by the Korea government (MSIT, RS-2025-00516062). This research was supported by funding from Korean government (Korea AeroSpace Administration (KASA), grant number RS-2026-25587698). Y. K. was supported by the National Research Foundation of Korea (NRF) grant funded by the Korean government (MSIT) (No. RS-2021-NR062087, RS-2026-25476464). JM gratefully acknowledges support from ANID MILENIO NCN2024\_112. SHOH acknowledges a support from the National Research Foundation of Korea (NRF) grant funded by the Korea government (Ministry of Science and ICT: MSIT) (No. RS-2026-25469119).

These observations were obtained by the James Clerk Maxwell Telescope, operated by the East Asian Observatory on behalf of The National Astronomical Observatory of Japan; Academia Sinica Institute of Astronomy and Astrophysics; the Korea Astronomy and Space Science Institute; the National Astronomical Research Institute of Thailand; Center for Astronomical Mega-Science (as well as the National Key R\&D Program of China with No. 2017YFA0402700). Additional funding support is provided by the Science and Technology Facilities Council of the United Kingdom and participating universities and organizations in the United Kingdom and Canada. Nāmakanui was constructed and funded by ASIAA in Taiwan, with funding for the mixers provided by ASIAA and at 230GHz by EAO. The Nāmakanui instrument is a backup receiver for the GLT. The authors wish to recognize and acknowledge the very significant cultural role and reverence that the summit of Maunakea has always had within the indigenous Hawaiian community.  We are most fortunate to have the opportunity to conduct observations from this mountain. We acknowledge the usage of the HyperLeda database (http://leda.univ-lyon1.fr). We acknowledge the use of NASA's SkyView facility (http://skyview.gsfc.nasa.gov) located at NASA Goddard Space Flight Center.
\end{acknowledgments}



%
\facilities{JCMT, MeerKAT, CFHT, GALEX}

\software{
Astropy \citep{astropy:2013, astropy:2018},   
Matplotlib \citep{Hunter:2007}, 
NumPy \citep{harris2020array}, 
SciPy \citep{2020SciPy-NMeth}
}




\bibliography{jcmt_vcc1249}{}
\bibliographystyle{aasjournalv7}



\end{document}